\documentclass[useAMS,usenatbib]{mn2e}
\usepackage{graphicx,subfig,amssymb}

\title[Globular Clusters in NGC 147 and NGC 185]
  {Newly-Discovered Globular Clusters in  NGC 147 and NGC 185 from PAndAS}
\author[Veljanoski et al.] {J.~Veljanoski$^1$,  A. M. N.~Ferguson$^1$,  A. P.~Huxor$^2$, A. D.~Mackey$^3$, C. K.~Fishlock$^3$,
\newauthor
M. J.~Irwin$^4$, N.~Tanvir$^5$, S. C.~Chapman$^4$, R. A. ~Ibata$^6$, G. F. ~Lewis$^7$
\newauthor
and A.~McConnachie$^8$ 
\\
$^1$ Institute for Astronomy, University of Edinburgh, Royal Observatory, Blackford Hill, Edinburgh, EH9 3HJ, UK\\
$^2$ Zentrum f\"{u}r Astronomie, Astronomisches Rechen-Institut M\"{o}nchhofstra{\ss}e  12-14, 69120 Heidelberg, Germany \\
$^3$ Research School of Astronomy $\&$ Astrophysics, Australian National University, Mt. Stromlo Observatory, Cotter Road,\\ Weston Creek, ACT 2611, Australia\\
$^4$ Institute of Astronomy, Madingley Road, Cambridge, CB3 0HA, UK \\
$^5$ Department of Physics $\&$ Astronomy, University of Leicester, Leicester LE1 7RH \\
$^6$ Observatoire de Strasbourg, 11, rue de l'Universit\'e, F-67000 Strasbourg, France \\
$^7$ Institute of Astronomy, School of Physics, University of Sydney, NSW 2006, Australia \\
$^8$ NRC Herzberg Institute of Astrophysics, 5071 West Saanich Road, Victoria, British Columbia V9E 2E7, Canada \\
\\}
\date{Released 2013 Xxxxx XX}

\pagerange{\pageref{firstpage}--\pageref{lastpage}} \pubyear{2013}

\def\LaTeX{L\kern-.36em\raise.3ex\hbox{a}\kern-.15em
    T\kern-.1667em\lower.7ex\hbox{E}\kern-.125emX}

\begin{document}

\label{firstpage}

\maketitle

\begin{abstract}
  Using data from the Pan-Andromeda Archaeological Survey (PAndAS), we
  have discovered four new globular clusters (GCs) associated with the
  M31 dwarf elliptical (dE) satellites NGC~147 and NGC~185. Three of
  these are associated with NGC~147 and one with NGC~185. All lie
  beyond the main optical boundaries of the galaxies and are the most
  remote clusters yet known in these systems. Radial velocities
  derived from low resolution spectra are used to argue that the GCs
  are bound to the dwarfs and are not part of the M31 halo population.
  Combining PAndAS with UKIRT/WFCAM data, we present the first
  homogeneous optical and near-IR photometry for the entire GC systems
  of these dEs.  Colour-colour plots and published colour-metallicity
  relations are employed to constrain GC ages and metallicities. It is
  demonstrated that the clusters are in general metal poor ([Fe/H] $<
  $ -1.25 dex), while the ages are more difficult to constrain. The
  mean (V-I)$_0$ colours of the two GC systems are very similar to
  those of the GC systems of dEs in the Virgo and Fornax clusters, as
  well as the extended halo GC population in M31.  The new clusters
  bring the GC specific frequency ($S_{N}$) to $\sim9$ in NGC~147 and
  $\sim5$ in NGC~185, consistent with values found for dEs of similar
  luminosity residing in a range of environments.
\end{abstract}

\begin{keywords}
 galaxies: globular glusters - galaxies: individual (NGC 147) - galaxies: individual (NGC 185)
\end{keywords}

\section{Introduction}

Globular clusters (GCs) are ubiquitous in massive galaxies and their
properties are believed to contain important clues about the galaxy
assembly process.  GCs are thought to form during major star forming
episodes which occur in the early Universe, as well as in subsequent
major merger events \citep[e.g.][]{West04,GCReview06}.  Their relative
ages and metallicities therefore help constrain the star formation
timescale of the host galaxies in which they reside.  GCs can also be
accreted onto galaxies alongside their hosts in minor mergers. At
present, we witness the Sagittarius dwarf spheroidal galaxy being
accreted onto the Milky Way (MW) while donating at least five GCs to
the MW GC system \citep[e.g.][]{Bellazzini03,Law10}. Furthermore,
almost 80\% of the outer halo GCs in M31 lie on top of stellar streams
supporting an accretion origin for most of them \citep{Mackey10b}.
Studying the GC systems of satellite galaxies is not just interesting
in its own right but can also give insight into how the GC systems of
more massive galaxies are assembled.

Detailed study of GCs in satellite galaxies is uniquely accessible
within the Local Group (LG) since individual stars can be easily
resolved from both space and the ground \citep[e.g.][]
{MackeyHST06,mackey10a,Dotter11}. Identifying LG GCs does not need to
rely on colour and magnitude criteria alone, as almost any globular
cluster candidate can be visually confirmed, effectively removing the
possibility of contamination.  Two particularly interesting LG
satellites are the dwarf ellipticals (dEs) NGC 147 and NGC 185.
Amongst the brightest of the LG dwarf galaxies, they reside in the
outer halo of M31 at projected galactocentric radii of $\sim100$ kpc,
and 3D distances of 118$^{+15}_{-15}$ kpc and 181$^{+25}_{-20}$ kpc
respectively \citep{conn12}. For reference, taking the halo core
radius of M31 to be $\sim$ 5 kpc \citep{Gilbert12}, the 3D distances
of NGC 147 and NGC 185 correspond to $\sim 24$ and $\sim 36$ core 
radii respectively. It has been argued that the two systems may
form a physical binary \citep{vdBergh98,Geha10} although this claim
has been questioned \citep{BattinelliDemers04,Watkins13}. The stellar
populations and star formation histories of these two dwarf galaxies
have been extensively studied
\citep[e.g.][]{LeeFreedmanMadore93,YoungLo97,Butler05,McConnachie05,Davidge05}.
Despite their initial similarities, they exhibit some notable
differences. NGC 147 is a typical dE composed primarily of old stars
and is dust and gas free \citep{YoungLo97}. However, NGC 185 has a
substantial number of intermediate age stars
\citep{MartinezDelgado99}, as well as gas and dust \citep{Young01}.
\citet{Bender91} found NGC 147 to be rotating, while NGC 185 to be
entirely pressure supported. In contrast, \citet{Geha10} found both
galaxies to have significant internal rotation through study of stars
out to 8 effective radii.

Previous studies have discovered a number of GCs in NGC 147 and NGC
185, all of which lie in or near the main bodies of these systems
\citep{Baade44,Hodge76,FJJ77,SharinaDavoust09}. Thorough photometric
\citep{Baade44,Hodge74} and spectroscopic studies
\citep{daCostaMould88,sharina06ages,SharinaDavoust09} have been
undertaken on these clusters, some of which are the brightest GCs
known to reside in LG dwarf galaxies. This work has shown that in
general the GCs hosted by NGC 147 and NGC 185 are old ($>$7 Gyrs) and
metal poor ([Fe/H] $<$ -0.8 dex).  However, it is becoming
increasingly appreciated that GCs can reside far from their host
galaxies \citep[e.g.][]{GalletiRBC04,hwang11,Huxor11,huxor12,jang12}
and this has motivated us to explore the remote environs of these
systems to search for additional members. In this paper, we present
the discovery of four outlying GCs, three of which belong to NGC 147
while one is hosted by NGC 185.  We derive the first homogeneous
optical and near-IR photometry for the entire GC populations of the
dEs which we use to constrain their ages and metallicities. In
addition, we present radial velocity measurements for the
newly-discovered GCs which are used to argue that they are bound to
the dEs themselves as opposed to the extended M31 halo system.

For the purposes of this work, we adopt distance moduli of 24.26$\pm$
0.06 and 23.96$^{+0.07}_{-0.06}$ for NGC 147 and NGC 185 respectively
\citep{conn12}. The absolute magnitudes of these dwarf galaxies are
taken from \citet{McConnachie12} and are -14.6$\pm$ 0.1 and -14.8$\pm$
0.1 for NGC 147 and NGC 185 respectively.

The paper has the following structure. The observations and data
reduction are described in Section 2.  Sections 3 and 4 present the
discovery and the photometric data of all known GCs in these two dwarf
galaxies.  The radial velocities of the newly-discovered GCs are
presented in Section 5. In Section 6, the optical and near-IR colours
are used to constrain ages and metallicities and we compare some
properties of the GCs to those observed in other systems. The content
of the paper is summarised in Section 7. Finally, in Appendix A we
present a review of the discovery history of the GCs found around NGC
147 and NGC 185, pointing out certain inconsistencies in the
literature, with the aim being to reduce the possibility of confusion
when studying these systems in the future.

\section{Observations and data reduction}

\subsection{Optical data}

The optical imaging used in this work was taken as part of the
recently completed ``Pan-Andromeda Archaeological Survey"
\citep[PAndAS;][]{mcconnachie09}. This optical imaging survey mapped
M31 and its close companion M33 with the wide-field MegaCam camera
\citep{Boulade03} on the Canada-France-Hawaii Telescope (CFHT).
In brief, the detector consists of a mosaic of 36 CCDs, giving a total
field of view of $\approx$ 1$\deg$ square, and with a pixel scale of
0.187 arcsec.  The survey consisted of $\approx$~400 distinct
pointings, covering an area of $\sim$380 square degrees and extending
to a projected radius of $\sim$150 kpc from the centre of M31. PAndAS
was undertaken in the \emph{g} and \emph{i} bands. The observations
were taken in good photometric conditions, with typical seeing $<$
0.8\arcsec, and reaching a depth of \emph{g} $\sim$ 25.5 and \emph{i}
$\sim$ 24.5 with S/N of 10.

The MegaCam data were initially reduced by CFHT staff using the
\emph{``Elixir''} pipeline which performs the standard bias,
flatfield, and fringe corrections, and determines the photometric
zero-point. The typical night-to-night variation of the zero-point is
around 1-2\% \citep{Regnault09}. The data are further processed by the
Cambridge Astronomy Survey Unit\footnote{http://casu.ast.cam.ac.uk},
as described in full detail by \citet{McConnachie10}.

\subsection{Near-IR data}

The near-IR data were taken in October 2008, as part of a survey
designed to look at red stellar populations in Local Group galaxies.
It used the Wide-Field Camera (WFCAM, \citet{Casali07}) on the
United Kingdom Infrared Telescope (UKIRT). This instrument has a pixel
scale of 0.4 arcsec, and the detectors are arranged such that four
dithered pointings are aligned to cover a square of 0.75 deg$^{2}$.
The observations were done in three near-IR bands, J, H and K, with
seeing of 0.8\arcsec\ or better, using the microstepping option to
improve the pixel sampling to 0.2\arcsec.

The data were reduced with a pipeline designed by the Cambridge
Astronomy Survey Unit, performing the usual dark-correction,
flatfielding, crosstalk removal, systematic noise and sky removal.
The pipeline, which is part of the VDFS\footnote{VISTA data flow
  system}, also does full astrometric and photometric calibration
based on the 2MASS point source catalogue and is described in
\citet{Cioni08} and \citet{Hodgkin09}. The nightly zero-point
variation on photometric nights is $<1$\%. The reduced images were
stacked and microstepped to produce individual detector frames. These
were then resampled to form a 0.333 arcsec per pixel science mosaic
\citep{Irwin04}.

\subsection{Spectroscopic data}

Low resolution spectra of the new GCs were obtained during two nights
in September 2010 using the ISIS instrument mounted on the William
Herschel Telescope (WHT). Several exposures were made of each object
with varying integration time depending on the brightness of the
target. The specifics of these observations are shown in Table
\ref{tab:specObs}.

\begin{table}
 \caption{Journal of the spectroscopic observations.}
 \label{tab:specObs}
 \begin{tabular}{cccc}
 	\hline
	\hline
	ID			& Date of obs.		& Number of 		& Integration time\\
				&			& exposures 		& per exposure [s] \\
	\hline
	PA-N147-1		& 10/09/2010	 	& 4			& 800 \\
	PA-N147-2		& 10/09/2010		& 3			& 800 \\
	PA-N147-3 		& 10/09/2010 		& 3			& 1500 \\
	\hline
	PA-N185			& 11/09/2010		& 4			& 1800 \\
	\hline
 \end{tabular}
\end{table}

The ISIS instrument employs two detectors, each attached to a separate
`arm' of the spectrograph.  The blue arm used the R600B grating and
was selected to cover the wavelength range between 3500 - 5100 \AA\
with a dispersion of 0.45 \AA/pix. The the red arm used the R600R
grating and was selected to cover the wavelength range between 7400 -
9100 \AA\ with a dispersion of 0.79 \AA/pix. The slit width was
2\arcsec\ throughout the observations and no binning was applied in
either the spectral or spatial directions. The S/N per pixel of the
spectra observed with the blue arm vary between 7 and 22, while for
the red arm between 11 and 30.  The observations were undertaken under
generally good atmospheric conditions with the seeing varying from 0.7
to 1.6\arcsec\ over the two nights.

The reduction of the spectra (bias subtraction, flat-fielding,
illumination correction) was done with standard tasks that are part of
the \textsc{ccdred} package within IRAF\footnote{IRAF is distributed
  by the National Optical Astronomy Observatories, which are operated
  by the Association of Universities for Research in Astronomy, Inc.,
  under cooperative agreement with the National Science Foundation}.
Following initial processing, the spectra were traced and extracted
with a $4^{\prime\prime}$ aperture using the \emph{apall} task in
IRAF.  For each exposure, a background region was interactively
selected and the sky was fit and subtracted with a second-order
Chebyshev polynomial.  The sky-subtraction generally worked well, with
only the strongest lines in the red spectra leaving some residuals.
The spectra were traced using a 3rd order cubic-spline function and
extracted with the optimal variance weighting option of \emph{apall}.

\begin{figure*}
\begin{center}
  \subfloat{\includegraphics[width=175mm,angle=0,trim=3mm 0mm 0mm 0mm,clip] {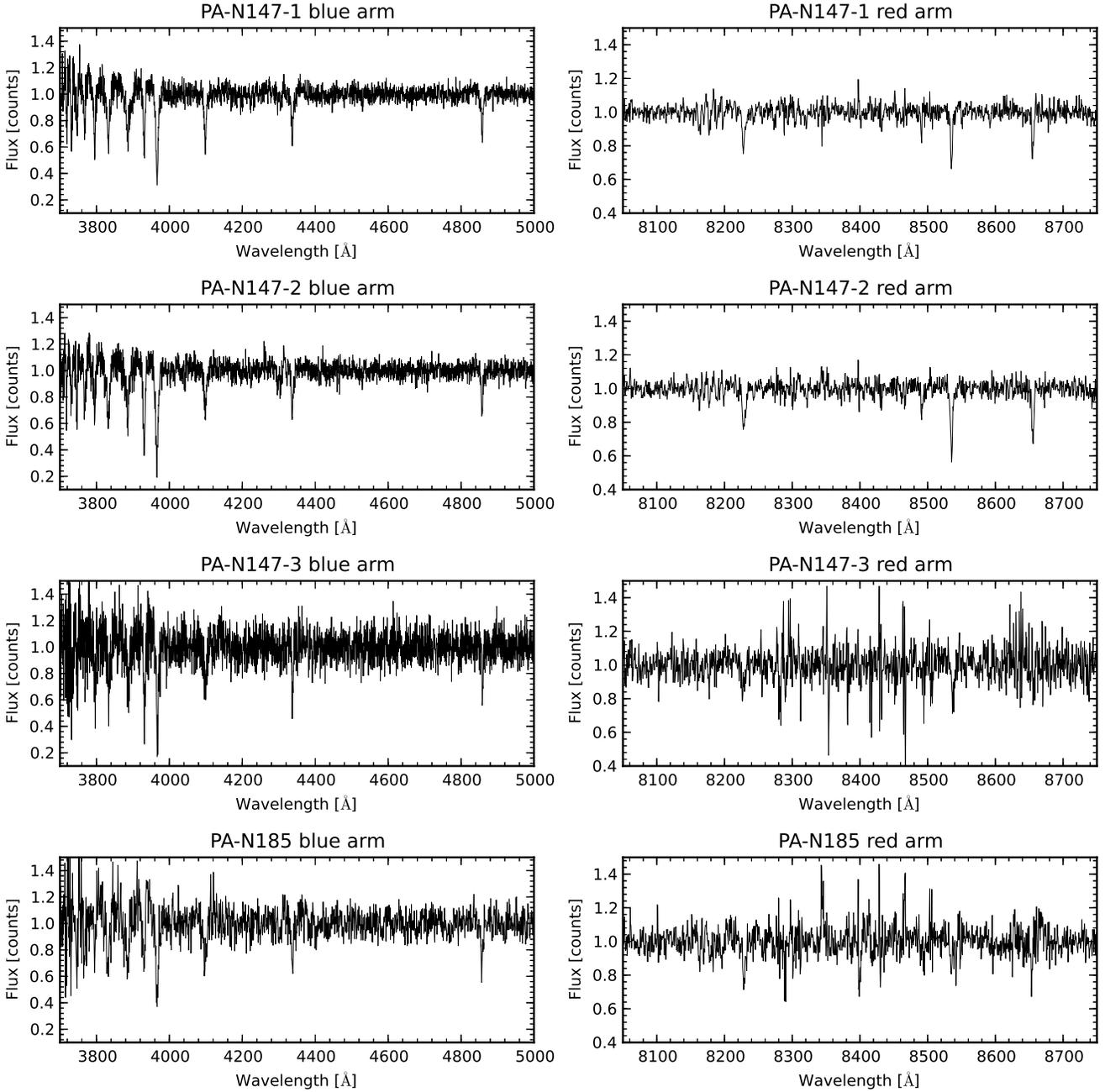}}     
  \caption{Wavelength calibrated, normalised spectra of the
      newly-discovered GCs.  The left (right) panels show the spectra
      obtained with the blue (red) arm of the ISIS instrument.}
\label{fig:sampleSpectra}
\end{center}
\end{figure*}

Wavelength calibration of the 1D spectra was based on He-Ne-Ar lamp
exposures obtained before and after each target exposure. The arc
spectra were extracted using the same \emph{apall} settings as the
target objects that they are used to calibrate. Typically $\sim$90
blue and $\sim$25 red lines were identified and the dispersion
solution was fit with a 3rd order cubic-spline function.  The RMS
residuals of these fits were 0.05$\pm$0.01 \AA\ in the blue and
0.02$\pm$0.01 \AA\ in the red. The IRAF \emph{dispcor} task was then
used to assign the wavelength solutions to the target GC spectra.  The
positions of sky lines were used to verify that the wavelength
calibration was accurate to 0.08 \AA\, with no systematic shifts. The
wavelength calibrated 1D spectra for each object were stacked using
inverse variance weights to produce final science spectra. Finally,
the continuum-normalised spectra are shown in Figure
\ref{fig:sampleSpectra}.

\section{Discovery}

\begin{figure}
\centering
  \includegraphics[width=93mm,height = 175mm,trim= 11mm 23mm  0mm 20mm,clip] {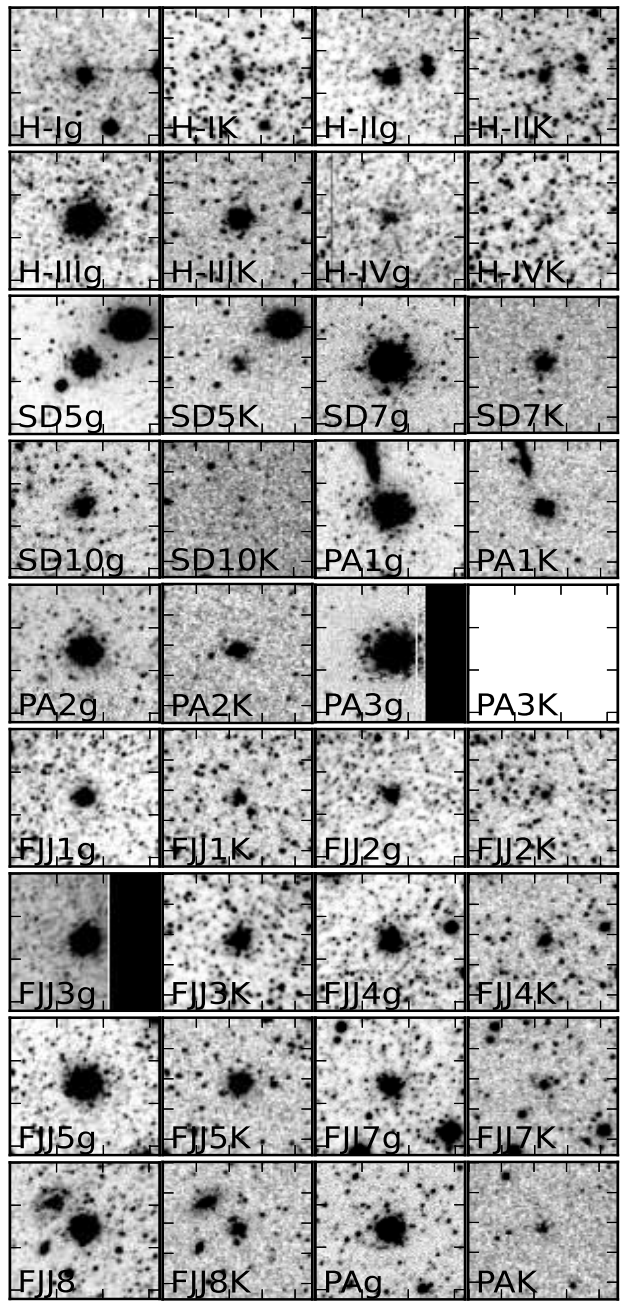} \\
  \caption{$g$ and K band images of all the clusters in NGC 147 and
    NGC 185. Each image is 30 $\times$ 30\arcsec wide. North is up and
    east is left. Cluster PA-N147-3 falls outside the IR survey hence no
    K-band image is shown. Note that the scale in each cutout has been
    manually set to make each cluster easily visible.   } 
  \label{fig:cutouts}
\end{figure}

At the distance of NGC 147 and NGC 185, GCs are partially resolved
into stars at optical wavelengths and can be easily identified.
Visual inspection of the area surrounding the two dEs was conducted by
three of us (A. Huxor, D. Mackey, J. Veljanoski) and resulted in the
discovery of four new GCs, three belonging to NGC 147, and one to NGC
185.  We refer to these objects with the prefix `PA'.  Images of these
objects in the \emph{g} and K bands are shown in Figure
\ref{fig:cutouts}. For completeness and comparison, the
previously-known GCs are also included. Table \ref{tab:147coords}
lists the coordinates, projected galactocentric radii (R$_{\rm proj}$)
and Galactic reddening coefficients, \emph{E}(B-V), from
\citet{Schlegel98} for all clusters.  While this paper was in
preparation, we found various inconsistencies in the literature
regarding the nomenclature of the previously-known GCs in NGC 147 and
NGC 185. These have been described in Appendix~A and we have made an
effort to rectify these inconsistencies throughout the paper whenever
possible.

Our search is based on homogeneous and sensitive imagery covering a
very large fraction of the areas around these two dEs. However,
there are two sources of incompleteness that need to be considered.
One is due to spatial incompleteness, arising from the many gaps
between the imaging CCDs in the MegaCam focal plane, and the gaps
coming from the imperfect tiling of the observations. GCs are
sufficiently small on the sky that they could be missed were they to
land in such a gap. The other source of incompleteness is due to our
ability to visually detect GCs. We assessed incompleteness using the
methodology we devised for our M31 halo GC search, which will be
discussed in full detail in Huxor et al. (2013, in prep).  Here we
provide only a brief overview.

To quantify the spatial incompleteness, we calculate the area lost
due to gaps in the imagery.  First, we used the WCS information in
the image headers to determine the RA and Dec boundaries of each of
the 36 CCDs in a given MegaCam pointing. We then generated a set of
random points around the two galaxies with a density of $\sim 100$
arcmin$^{-2}$.  The $"$observed" area was simply calculated by
dividing the number of points which fell within the RA and Dec
boundaries by the total number generated. This test indicates that
the fractions of NGC 147 and NGC 185 that are imaged out to a
circular radius of 15 kpc are 96.3\% and 93.5\% respectively.

To quantify our ability to detect GCs through visual inspection, we
generated a suite of artificial GCs using the {\sc SIMCLUST}
\citep{Deveikis08} and {\sc SKYMAKER} \citep{Bertin09} software
packages. The clusters were generated with a range of luminosity and
concentration\footnote{The concentration is defined as $c = {\rm 
log}(r_t/r_c)$; where $r_t$ and $r_c$ are the cluster tidal and
core radii respectively, assuming a \citet{King62} profile for its
radial surface density.} and were embedded within field star
backgrounds of varying density.  Thumbnails of these generated
clusters were then visually inspected by one of us (A. Huxor) who
decided if a cluster was present or not based on the same criteria
we applied to the real data.  Using this method, we found that, for
most values of concentration, the GC recovery is 100(50)\% complete
down to $M_{V} \approx -5.3(-4.1)$ around NGC 147.  Around NGC 185,
the search is 100(50)\% complete down to $M_{V} \approx -5.0(-3.8)$.
For reference, these limits indicate that we would be able to detect
most, if not all, of the Milky Way Palomar type clusters. Indeed, of
the thirteen such objects known in the Milky Way that have available
absolute integrated V magnitudes, ten of them are brighter than
-5.0, while eleven are brighter than -4.0 \citep{Harris96Cat}.  We
would not be able to detect the Koposov clusters which have $M_V
\sim$-1 \citep{Koposov07}, although these are difficult to find even
in the Milky Way. It thus seems likely that we have uncovered all
the luminous remote GCs around these two systems.

\begin{table}
  \caption{Coordinates, projected radii and Galactic reddening coefficients of GCs in NGC 147 and NGC 185.}
 \label{tab:147coords}
 \begin{tabular}{ccccc}
 	\hline
	\hline
	ID			& RA(J2000)		& DEC(J2000)		& R$_{\rm proj}$	&\emph{E}(B-V)	\\
				& [h  m  s]		& [d  m  s]		& [kpc]			&[mag]	\\
	\hline	
	Hodge I			& 00 33 12.2		& +48 30 32.3		& 0.03			&0.17	\\
	Hodge II		& 00 33 13.6		& +48 28 48.7		& 0.34			&0.17	\\
	Hodge III		& 00 33 15.2		& +48 27 23.1		& 0.64			&0.17	\\
	Hodge IV		& 00 33 15.0		& +48 32 09.6		& 0.38			&0.17	\\
	SD-GC5			& 00 32 22.9		& +48 25 49.0		& 1.93			&0.18	\\
	SD-GC7			& 00 32 22.2		& +48 31 27.0		& 1.72			&0.17	\\
	SD-GC10			& 00 32 47.2		& +48 32 10.7		& 0.92			&0.17	\\
	PA-N147-1		& 00 32 35.3	 	& +48 19 48.0		& 2.53			&0.15	\\
	PA-N147-2		& 00 33 43.3		& +48 38 45.0		& 2.04			&0.16	\\
	PA-N147-3	 	& 00 34 10.0  		& +49 02 39.0		& 6.97			&0.16	\\
\hline
	FJJ I			& 00 38 42.7		& +48 18 40.4		& 0.53			&0.17	\\
	FJJ II			& 00 38 48.1   		& +48 18 15.9		& 0.45			&0.17	\\
	FJJ III			& 00 39 03.8    	& +48 19 57.5		& 0.20			&0.19	\\
	FJJ IV			& 00 39 12.2		& +48 22 48.2		& 0.64			&0.19	\\
	FJJ V			& 00 39 13.4    	& +48 23 04.9		& 0.70			&0.19	\\
	FJJ VII 		& 00 39 18.4 		& +48 23 03.6		& 0.81			&0.19	\\
	FJJ VIII		& 00 39 23.7		& +48 18 45.1		& 0.83			&0.17	\\
	PA-N185			& 00 38 18.8		& +48 22 04.0		& 1.20			&0.18	\\
	\hline

 \end{tabular}
\end{table}

\begin{figure}
\centering
  \subfloat{\includegraphics[width=85mm,angle=270,trim=0 0mm 0 0,clip] {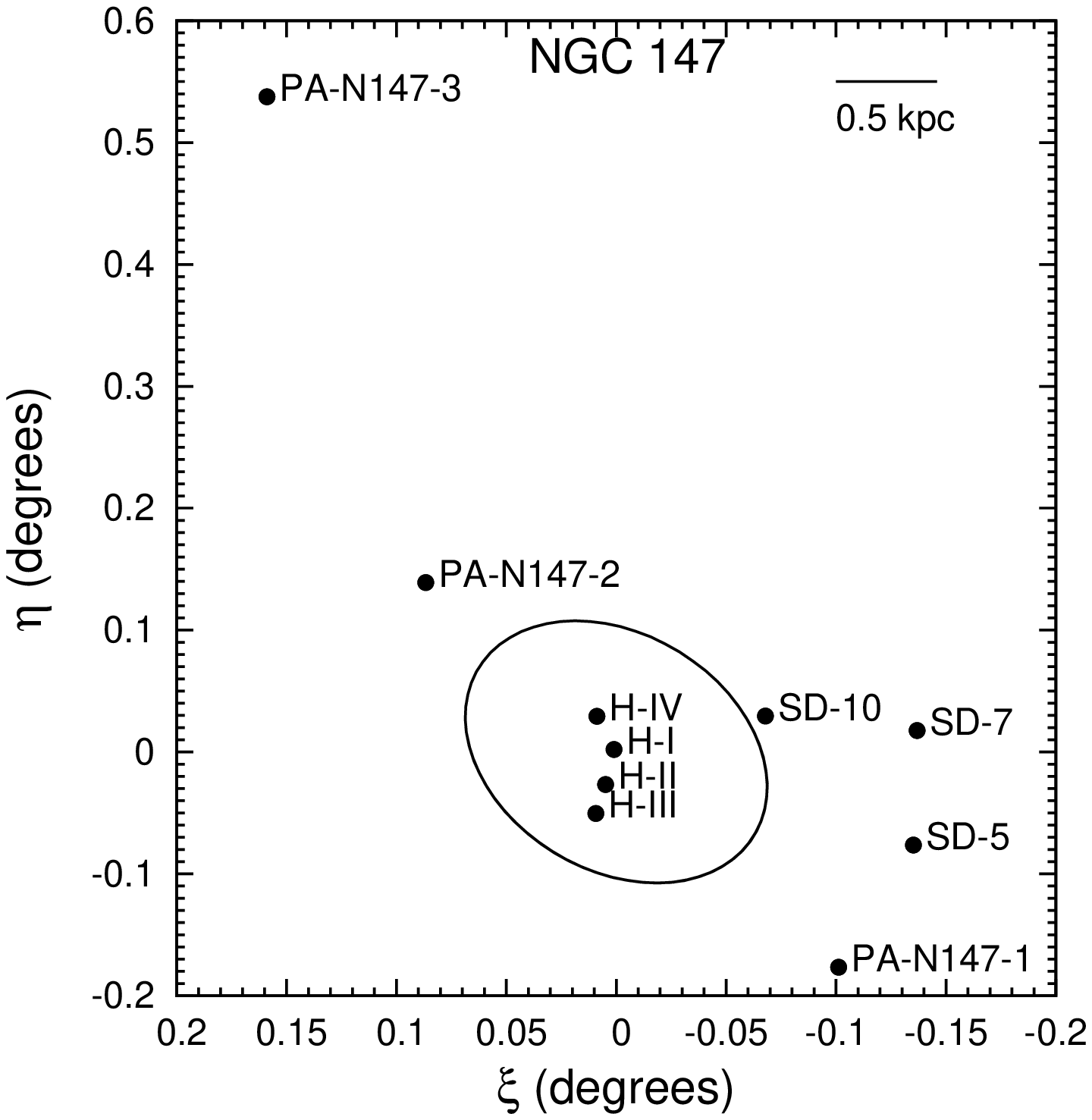}} \\               
  \subfloat{\includegraphics[width=85mm,angle=270,trim=0 0mm 0 0,clip] {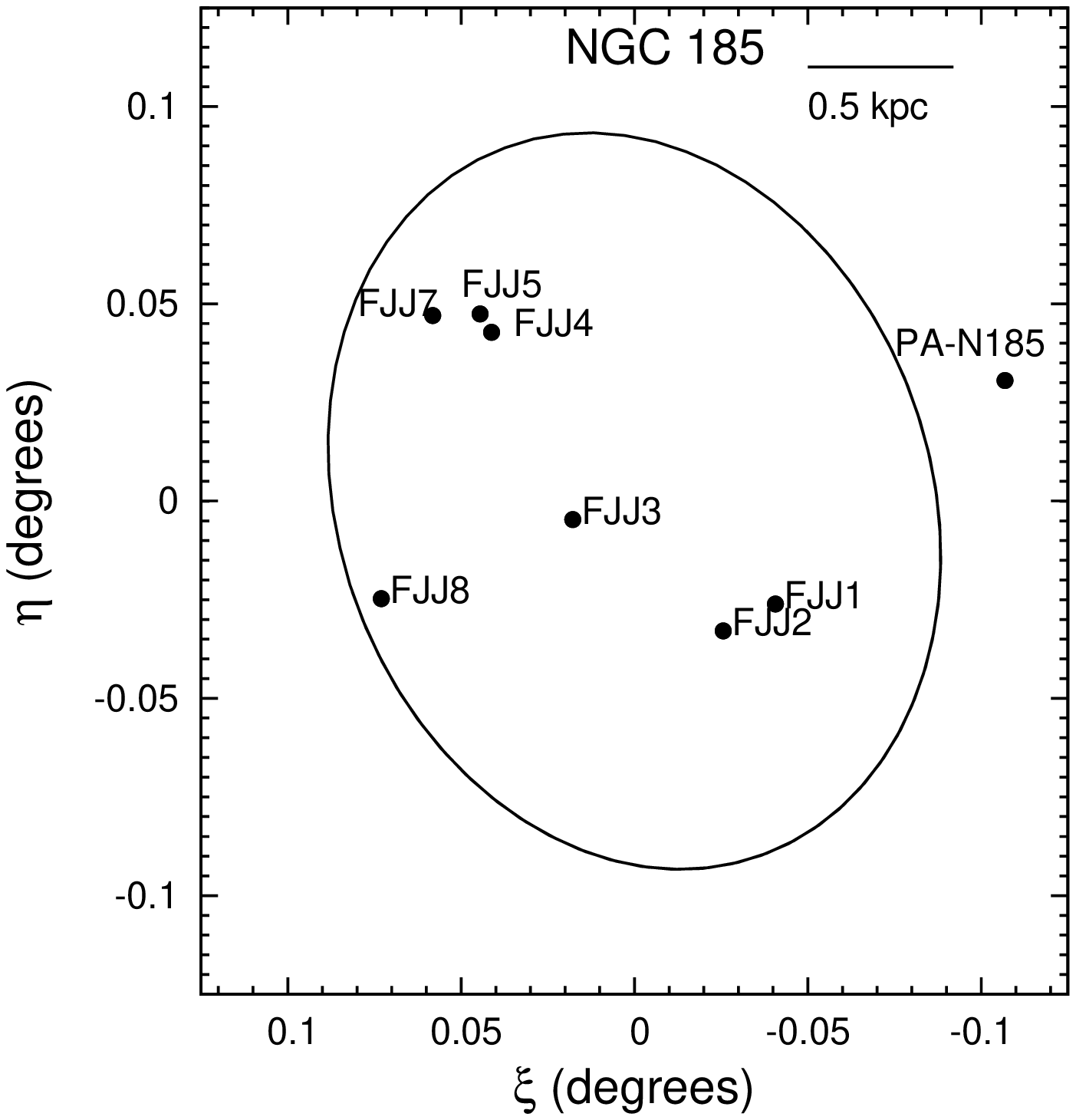}}                
  \caption{Schematic view of NGC 147 (top panel) and NGC 185
      (bottom panel) along with their GCs in standard coordinates. The
      centres of NGC 147 and NGC 185 are taken to be RA=00:33:12.1,
      Dec= 48:30:32 and RA=00:38:58.0, Dec=48:20:15 respectively 
      \citep{McConnachie12}. The ellipses represent the 25th magnitude 
      B-band isophote taken from \citet{RC3b}. The new GCs lie further 
      away from the centre of their host galaxies than the known populations. 
      On the top panel ``H" stands for ``Hodge" and ``SD" stands for ``GC-SD".}
  \label{fig:spatial}
\end{figure}

Figure \ref{fig:spatial} shows the spatial distribution of GCs in the
two galaxies where it can readily be seen that the newly-discovered
GCs are much more remote than the known populations. Given this, it is
possible that some of them might be bound to M31 rather than to NGC
147 or NGC 185. This could be especially true for PA-N147-3 which is
located at projected radius of 6.6 kpc from the centre of its host
galaxy NGC 147.

To examine this possibility, we calculate the probability of finding a
genuine M31 GC projecting as close to NGC 147 and NGC 185 as the new
objects we have found. We consider a circular annulus spanning 90--110
kpc in projected radius from M31's centre, the size of which is chosen
to comfortably encompass the positions and extents of NGC 147 and NGC
185. Inside the annulus there are 8 M31 GCs excluding the new objects
around NGC 147 and NGC 185 (Mackey et al. 2013, in prep.), which in
turn equals a number density of $6.4\times 10^{-4}$
clusters/kpc$^{2}$.  We define a search radius of 7 kpc and 2 kpc
around NGC 147 and NGC 185 respectively, chosen to be slightly larger
than the GCs that have the largest projected distance from the centres
of these systems. On average one would expect to find 0.008 GCs within
a circular area of 2 kpc radius and 0.098 GCs within a 7 kpc circular
radius.  The Poisson probability of finding one or more M31 GCs within
a 2 kpc radius of NGC 185 and within a 7 kpc radius of NGC 147 is 1\%
and 10\% respectively.  The probability of finding three or more GCs
within a 7 kpc radius of NGC 147 drops to 0.01\%.  The conclusion from
this analysis is that, in absence of other information, there is a
small chance that at least one of the newly-discovered GCs around NGC
147 could be part of the M31 halo GC system, while the newly
discovered cluster around NGC 185 almost certainly belongs to its
apparent host galaxy. It is also worth noting that two of the
newly discovered GCs around NGC 147 appear to lie on top of tidal
tails which emanate from this system \citep[][Irwin et al. in
prep]{Lewis13}.

\section{Photometry}

Integrated photometry of all the GCs in these two dwarf galaxies was
done with the \emph{phot} task within IRAF. The centroid algorithm
within the \emph{phot} task was used to determine accurate centres of
the clusters by computing the intensity weighted means of the marginal
profiles along the physical x and y axes in the g-band images.
Circular apertures were then used to sum the total flux.  Concentric
``sky'' annuli around the photometric apertures, with typical width of
$\sim 7^{\prime\prime}$, were employed to determine the mean
background flux contribution. The exact sizes of the background annuli
were carefully selected to avoid pollutants such as bright foreground
stars and background galaxies. As some of these clusters lie in very
crowded fields, detailed tests were performed to ensure that the
background contribution was reliably subtracted.

For each GC we constructed a curve-of-growth by measuring the flux in
concentric apertures of increasing radius. The final photometric
aperture size was set to be the point at which the cumulative flux
becomes flat. This did not work for clusters in the very dense fields,
where we adopted an aperture size based mainly on visual inspection,
even though the curve-of-growth was not flat. To reduce background
contamination, a smaller aperture of $3.5^ {\prime\prime}$ was used
when measuring colours, valid if we assume that no colour gradients
are present within the GCs. Aperture sizes for both colours and
magnitudes were matched to be the same in the optical and near-IR
data. For all photometric measurements, the instrumental magnitudes
were zero-point calibrated, corrected for atmospheric extinction and
corrected for Galactic reddening using the \citet{Schlegel98} maps
interpolated to the position of each GC.  Tables \ref{tab:mags} and
\ref{tab:colors} list the extinction-corrected magnitudes and colours
of all GCs in NGC 147 and NGC 185.

Some objects required special treatment. In the cases of Hodge II and
SD-GC5, a mask was applied consistently to all filters in order to
exclude bright contaminating objects that entered the photometric
aperture. The cluster PA-N147-3 was split between two separate CCD
frames in the optical data. To account for this, we summed the flux
within rectangular apertures on each frame and combined them; this
strictly represents a lower limit on the total magnitude of this
object. This same cluster has the largest projected radius of all GCs
in our sample and it falls outside the near-IR survey area, meaning no
near-IR measurement is possible. The cluster SD-10 was not detected in
the near-IR imaging and the corresponding near-IR magnitudes listed
provide upper limits on its brightness. The GC FJJ III lies very close
to the edge of the chip in the \emph{g} band, while it is split
between two chips in the \emph{i} band. For the optical photometry of
this cluster, previously published data obtained with the Isaac Newton
Telescope Wide Field Camera (INT/WFC) were used \citep{McConnachie05}.
This was not possible for PA-N147-3 as it falls outside the area
covered by the INT survey. The INT/WFC survey uses the Johnson V and
Gunn \emph{i} passbands, which are on the Vega scale. Hence the
magnitudes of FJJ III were converted to the CFHT/MegaCam system, which
is on the AB scale, via transformations presented in \citet{Ibata07}.

\begin{table*}
 \centering
 \caption{ Photometry for the NGC147 and NGC 185 GCs. $M_{V,0}$ was
   calculated assuming the line-of-sight distance to the clusters is
   same as to their host galaxies. Superscripts refer to (1) the
   presence of masked bright objects in the photometric aperture, (2)
   the magnitude is a lower limit as the cluster is split between CCD
   frames and was measured in a rectangular aperture, (3) the
   photometry is measured from INT imaging. }
 \label{tab:mags}
 \begin{tabular}{@{}cccccccccc@{}}
 	\hline
	\hline
	ID		& Aperture	& $g_0$		& $i_0$		&$J_0$		& $H_0$		& $K_0$		& $V_0$		& $I_0$		& $M_{V,0}$ \\
			& [arcsecs]	 &[mag]	&[mag]	&[mag]	&[mag]	&[mag]	&[mag]	&[mag]	&[mag]	\\
	\hline
	 Hodge I 	&6.5	&17.17 $\pm$ 0.04		&16.31 $\pm$ 0.02		&14.84 $\pm$ 0.02    &14.25 $\pm$ 0.02 	&14.13 $\pm$ 0.03	&16.86 $\pm$ 0.03		&15.92 $\pm$ 0.02		& -7.40		\\
	 Hodge II$^1$ 	&4.6	&17.79 $\pm$ 0.04		&17.07 $\pm$ 0.04		&15.93 $\pm$ 0.02    &15.30 $\pm$ 0.02	&15.14 $\pm$ 0.02	&17.53 $\pm$ 0.03		&16.71 $\pm$ 0.04		& -6.73		\\
	 Hodge III 	&7.4	&16.29 $\pm$ 0.03		&15.65 $\pm$ 0.02		&14.71 $\pm$ 0.01    &14.15 $\pm$ 0.01	&14.18 $\pm$ 0.01	&16.05 $\pm$ 0.02		&15.29 $\pm$ 0.02		& -8.21		\\
	 Hodge IV	&4.6	&18.83 $\pm$ 0.04		&17.81 $\pm$ 0.03		&16.55 $\pm$ 0.03    &15.98 $\pm$ 0.03	&15.94 $\pm$ 0.04	&18.48 $\pm$ 0.03		&17.41 $\pm$ 0.03		& -5.78		\\
	 GC-SD5$^1$ 	&5.6	&17.83 $\pm$ 0.02		&17.20 $\pm$ 0.01		&16.36 $\pm$ 0.02    &15.81 $\pm$ 0.02	&15.77 $\pm$ 0.03	&17.60 $\pm$ 0.01		&16.85 $\pm$ 0.01		& -6.66		\\
	 GC-SD7 	&7.4	&16.73 $\pm$ 0.01		&16.03 $\pm$ 0.01		&15.03 $\pm$ 0.01    &14.49 $\pm$ 0.01	&14.43 $\pm$ 0.02	&16.47 $\pm$ 0.01		&15.66 $\pm$ 0.01		& -7.79		\\
	 GC-SD10	&4.6	&19.59 $\pm$ 0.01		&18.86 $\pm$ 0.01		&$>$17.84 $\pm$ 0.06 &$>$17.30 $\pm$ 0.06	&$>$17.22 $\pm$ 0.12	&19.32 $\pm$ 0.01		&18.49 $\pm$ 0.02	& -4.94		\\
	 PA-N147-1 	&6.5	&16.72 $\pm$ 0.01		&16.09 $\pm$ 0.01		&15.15 $\pm$ 0.01    &14.69 $\pm$ 0.01	&14.64 $\pm$ 0.02	&16.49 $\pm$ 0.01		&15.74 $\pm$ 0.01		& -7.77		\\
	 PA-N147-2 	&6.5	&17.11 $\pm$ 0.01		&16.44 $\pm$ 0.01		&15.37 $\pm$ 0.01    &14.89 $\pm$ 0.01	&14.86 $\pm$ 0.02	&16.87 $\pm$ 0.01		&16.08 $\pm$ 0.01		& -7.39		\\														                          
	 PA-N147-3 	& 	&17.17 $\pm$ 0.01$^2$		&17.63 $\pm$ 0.01$^2$		&...		     &...		&...			&17.38$\pm$ 0.01$^2$		&17.42 $\pm$ 0.01$^2$		& -6.88$^2$	\\
	 \hline
	 FJJ I 		&4.6	&17.96 $\pm$ 0.03		  &17.22 $\pm$ 0.02		&15.24 $\pm$ 0.02    &15.39 $\pm$ 0.03	&15.06 $\pm$ 0.02	&17.70 $\pm$ 0.03		&16.85 $\pm$ 0.03		& -6.26		\\
	 FJJ II 	&4.6	&18.28 $\pm$ 0.04		  &17.49 $\pm$ 0.03		&15.59 $\pm$ 0.03    &15.75 $\pm$ 0.03	&15.73 $\pm$ 0.04	&18.00 $\pm$ 0.03		&17.12 $\pm$ 0.03		& -5.96		\\
	 FJJ III	&7.4	&16.20 $\pm$ 0.29$^3$	&15.55 $\pm$ 0.11$^3$		&13.80 $\pm$ 0.02    &14.02 $\pm$ 0.02	&13.94 $\pm$ 0.01	&15.99 $\pm$ 0.17$^3$		&15.14 $\pm$ 0.11$^3$		& -7.97$^3$	\\
	 FJJ IV		&5.6	&17.58 $\pm$ 0.03		  &17.00 $\pm$ 0.02		&15.24 $\pm$ 0.02    &15.53 $\pm$ 0.03	&15.60 $\pm$ 0.04	&17.37 $\pm$ 0.02		&16.65 $\pm$ 0.02		& -6.59		\\
	 FJJ V		&7.4	&16.38 $\pm$ 0.03		  &15.66 $\pm$ 0.02		&13.85 $\pm$ 0.01    &14.09 $\pm$ 0.02	&14.00 $\pm$ 0.01	&16.12 $\pm$ 0.02		&15.30 $\pm$ 0.02		& -7.84		\\
	 FJJ VII	&4.6	&18.36 $\pm$ 0.02		  &17.62 $\pm$ 0.01		&15.80 $\pm$ 0.02    &16.06 $\pm$ 0.03	&16.07 $\pm$ 0.04	&18.10 $\pm$ 0.02		&17.25 $\pm$ 0.02		& -5.85		\\
	 FJJ VIII	&4.6	&17.29 $\pm$ 0.01		  &16.59 $\pm$ 0.01		&14.82 $\pm$ 0.01    &15.08 $\pm$ 0.02	&15.09 $\pm$ 0.02	&17.04 $\pm$ 0.01		&16.23 $\pm$ 0.01		& -6.92		\\
	 PA-N185	&4.6	&18.65 $\pm$ 0.01		  &17.98 $\pm$ 0.01		&16.23 $\pm$ 0.03    &16.54 $\pm$ 0.04	&16.39 $\pm$ 0.06	&18.41 $\pm$ 0.01		&17.62 $\pm$ 0.01		& -5.55		\\
	 
	\hline	

 \end{tabular}
\end{table*}

\begin{table*}
 \centering
 \caption{Colours for the NGC 147 and NGC 185 GCs. All measurements
   are done within a 3.5\arcsec aperture radius in order to minimise
   background contamination.  Colours of PA-N147-3 cannot be
   calculated as different portions of the cluster have been measured
   on the optical imaging and it was not imaged in the near-IR.
   Superscripts refer to (1) the presence of masked bright objects in
   the photometric apperture, (2) the photometry is measured from INT
   imaging. }
 \label{tab:colors}
 \begin{tabular}{@{}cccccccc@{}}
 	\hline
	\hline
	ID			& $(g-i)_0$			& $(g-J)_0$		& $(g-H)_0$		& $(g-K)_0$		& $(V-I)_0$			&$(V-K)_0$	\\
				&[mag]				&[mag]			&[mag]			&[mag]			&[mag]				&[mag]		\\
	\hline
	Hodge I			&0.72$\pm$0.04		&1.97$\pm$0.04	&2.51$\pm$0.04	&2.59$\pm$0.04	&0.83$\pm$0.04	&2.33$\pm$0.03	\\
	Hodge II$^1$		&0.68$\pm$0.07		&1.75$\pm$0.05	&2.30$\pm$0.05	&2.35$\pm$0.05	&0.79$\pm$0.07	&2.11$\pm$0.04	\\
	Hodge III		&0.61$\pm$0.04		&1.57$\pm$0.04	&2.07$\pm$0.04	&2.10$\pm$0.04	&0.73$\pm$0.04	&1.87$\pm$0.03	\\
	Hodge IV		&0.96$\pm$0.05		&2.31$\pm$0.05	&2.89$\pm$0.05	&2.96$\pm$0.05	&1.02$\pm$0.04	&2.62$\pm$0.05	\\
	GC-SD5			&0.60$\pm$0.02		&1.46$\pm$0.03	&2.00$\pm$0.02	&2.03$\pm$0.03	&0.73$\pm$0.01	&1.81$\pm$0.03	\\
	GC-SD7			&0.67$\pm$0.01		&1.69$\pm$0.02	&2.22$\pm$0.01	&2.28$\pm$0.02	&0.78$\pm$0.01	&2.04$\pm$0.02	\\
	GC-SD10			&0.76$\pm$0.02		&$<$1.83$\pm$0.06	&$<$2.40$\pm$0.05	&$<$2.25$\pm$0.11	&0.85$\pm$0.02	&$<$1.98$\pm$0.11	\\
	PA-N147-1		&0.52$\pm$0.01		&1.60$\pm$0.02	&2.04$\pm$0.01	&2.14$\pm$0.02	&0.75$\pm$0.01	&1.91$\pm$0.02	\\
	PA-N147-2		&0.66$\pm$0.01		&1.70$\pm$0.02	&2.16$\pm$0.02	&2.20$\pm$0.02	&0.79$\pm$0.01	&1.97$\pm$0.02	\\	
	PA-N147-3		&...				&...			&...			&...			&...			&...			\\
	\hline
	FJJ I			&0.73$\pm$0.04		&2.64$\pm$0.04	&2.41$\pm$0.04	&2.55$\pm$0.04	&0.84$\pm$0.04	&2.30$\pm$0.04	\\
	FJJ II			&0.75$\pm$0.05		&2.63$\pm$0.05	&2.46$\pm$0.05	&2.52$\pm$0.05	&0.85$\pm$0.04	&2.25$\pm$0.04	\\
	FJJ III$^2$		&0.67$\pm$0.31		&2.56$\pm$0.29	&2.30$\pm$0.29	&2.42$\pm$0.29	&0.86$\pm$0.19	&2.22$\pm$0.16	\\
	FJJ IV			&0.63$\pm$0.04		&2.39$\pm$0.03	&2.09$\pm$0.04	&2.10$\pm$0.04	&0.75$\pm$0.04	&1.87$\pm$0.04	\\
	FJJ V			&0.70$\pm$0.03		&2.52$\pm$0.03	&2.24$\pm$0.03	&2.31$\pm$0.03	&0.81$\pm$0.03	&2.06$\pm$0.02	\\
	FJJ VII			&0.73$\pm$0.03		&2.60$\pm$0.03	&2.34$\pm$0.03	&2.36$\pm$0.04	&0.84$\pm$0.03	&2.10$\pm$0.03	\\
	FJJ VIII		&0.69$\pm$0.02		&2.48$\pm$0.02	&2.22$\pm$0.03	&2.24$\pm$0.02	&0.80$\pm$0.02	&2.00$\pm$0.02	\\
	PA-N185			&0.68$\pm$0.01		&2.50$\pm$0.03	&2.22$\pm$0.03	&2.38$\pm$0.04	&0.80$\pm$0.01	&2.13$\pm$0.04	\\
	\hline	
 \end{tabular}
\end{table*}

To allow comparison with other work, Tables \ref{tab:mags} and
\ref{tab:colors} also contain magnitudes and colours converted to the
more widely used Johnson/Cousins filters. The CFHT/MegaCam data was
transformed into the standard V and I system via corrected relations
from \citet{Huxor08}\footnote{Note that the transformation equations
  in that paper were incorrectly written, but the magnitudes derived
  were based on the correct equations.}:

\begin{displaymath}
\begin{array}{l} 
g_1 = g + 0.092\\
i_1 = i - 0.401 \\
\textrm{V} = g_1 - 0.42(g_1 - i_1) + 0.04(g_1 - i_1)^2 + 0.10 \\
\textrm{I} = i_1 - 0.08(g_1 - i_1) + 0.06 \\
\end{array}
\end{displaymath} 
These relations were derived for the $i^\prime$ filter used with
CFHT/MegaCam pre-June 2007, while in October 2007 a new \emph{i}
filter was installed on this instrument. The data used in this paper
were taken with the new \emph{i} filter and so before transforming the
MegaCam data to the standard V and I filters, conversion from the new
\emph{i} to the old $i^\prime$ filter was done using the relation
derived in Ibata et al. (2013, in prep.).

In deriving magnitudes and colours, various uncertainties are included
and appropriately combined. The instrumental magnitude uncertainties
reported by IRAF are small, as the clusters are much brighter than the
background sky. For GCs lying within the main optical bodies
of their host dwarf galaxies, the main source of uncertainty comes
from the local background, which is contaminated by resolved stars
from the host galaxy as well as the underlying diffuse light.  To
assess the uncertainty in the background, we randomly placed 10
apertures around each cluster sampling the local sky. While the
apertures include the resolved field star component and the diffuse
light of the host dwarf galaxies, we excluded obvious contaminants
such as background galaxies or foreground Milky Way stars. The sky
apertures were chosen to have the same size as the magnitude and
colour apertures that we used to photometer the GCs. We found the
standard deviation of all 10 sky measurements around each cluster,
and added this in quadrature to the instrumental and zero-point
errors. Furthermore, every conversion between filters introduces
an additional uncertainty to the derived magnitudes and colours that 
we account for as well.

\section{Radial Velocities}

Heliocentric radial velocities of the newly-discovered GCs were
measured using a chi-squared minimisation technique between the GC
spectra and spectra of high signal-to-noise radial velocity template
stars and clusters \citep{Veljanoski13a}. This
method is analogous to the standard cross-correlation technique, and
produces similar results. The advantage of doing a chi-square
minimisation is that this technique uses the uncertainties in both the
template and target spectra, which helps to eliminate spurious
features, and differentiate between genuine spectral lines and poorly
subtracted sky lines, which might be important in the case of faint
targets.

Because both ISIS arms were used during the observations, two
independent velocity measurements could be made for each GC. The
values reported in Table \ref{tab:rv} represent the error-weighted
averages of the individual measurements from the blue and red spectra.
For comparison, the heliocentric radial velocities of NGC 147 and NGC
185 themselves are also shown \citep{Geha10}.

\begin{table}
  \caption{Heliocentric radial velocities and their uncertainties for the newly-discovered GCs 
    around NGC 147 and NGC 185. For comparison, the heliocentric velocities of NGC 147 and NGC 185 
    \citep{Geha10} are also shown.}
 \label{tab:rv}
 \begin{tabular}{ccc}
 	\hline
	\hline
	ID			& Radial velocity [km/s]	& Uncertainty [km/s] \\
	\hline
	NGC 147			& -193.1			& 0.8			\\
	PA-N147-1		& -215	 			& 10			\\
	PA-N147-2		& -219				& 10			\\
	PA-N147-3 		& -133				& 24			\\
	\hline
	NGC 185			& -203.8			& 1.1			\\
	PA-N185			& -254				& 15			\\
	\hline
 \end{tabular}
\end{table}

The measured velocities of the new GCs can be used as another
indicator of whether they are bound to the dEs or to M31. The velocity
dispersion of M31 halo GCs that lie beyond 70 kpc in projection is
$\sim$50 km/s \citep{Veljanoski13a}. Given the large difference
between the radial velocities of the new NGC 147 GCs and that of M31
(-301 $\pm$ 4 km/s, \citet{Courteau99}), it is likely that they are
hosted by the dwarf galaxy. This is not the case for PA-N185, as
the difference between it and the M31 velocity is comparable to the
M31 outer halo GC velocity dispersion. However, when combined with the
probability arguments presented earlier, the velocity measurements
strengthen the conclusion that the new GCs are probably members
of the dE systems and not the M31 halo.

It is interesting to point out that the cluster PA-N147-3 projects in
position halfway between NGC 147 and the newly-discovered dwarf
spheroidal Cass II (Conn et al. 2012, Irwin et al. in prep.).
Furthermore, it has a similar radial velocity to Cass II, which is
measured to be -145 $\pm$ 3 km/s \citep{Collins13}. It is therefore
possible that this GC could be a satellite of Cass II instead of NGC
147, or else not bound to either system.  We postpone a detailed
kinematical analysis of GCs within the NGC 147/185 subgroup until a
later publication.

\section{Analysis}
\subsection{Ages and Metallicities}

We derive age and metallicity estimates for the two cluster samples
using our integrated optical and NIR photometry.  In principle, one
can determine accurate ages and metallicities of GCs with high quality
spectroscopic data. However, our spectra have low signal-to-noise, and
only two out of the four new GCs have spectra that are suitable for
metallicity determination.  Furthermore, as our goal is to present a
homogeneous analysis of the sample, including objects for which we do
not have spectra, we prefer to base our analysis on integrated
photometry alone. Optical colours are well-known to suffer from an
age-metallicity degeneracy, however the addition of near-IR
measurements can greatly improve the situation
\citep[e.g.][]{Puzia02,Hempel05,Santos11}.  This is because of the
different sensitivities to age and metallicity of the optical-optical
and optical-near-IR colours. The optical $g$-band is most sensitive to
stars near the main sequence turn-off, the magnitudes of which are
mostly driven by age. Conversely, the near-IR K band is most sensitive
to red giant branch stars \citep{Saviane00,Yi01}. The $g$-$i$ and
$g$-K colours have similar sensitivity to age, but $g$-K measures the
temperature of the red giant branch which more reflects metallicity
than age. Plotting these colours on a colour-colour diagram and
comparing to simple stellar population (SSP) model tracks allow us to
derive estimates of the age and metal content of each GC.

Various SSP models have been constructed to date
\citep[e.g.][]{MarastonSSP98,BruzualCharlot03,MarigoPadova08,VazdekisSSP10}.
Despite improvements over time, discrepancies still exist between
models.  One of the largest intrinsic uncertainties in SSP models
comes from the limited understanding of certain phases of advanced
stellar evolution such as the thermally-pulsating asymptotic giant
branch (TP-AGB) phase. Different attempts to implement this phase have
led to large differences in the near-IR fluxes. Stars in this phase
are short-lived making calibration difficult \citep{GirardiPadova10}.
Models having prominent TP-AGB phases cause the near-IR luminosity of
objects to be overpredicted, but this only affects objects of young to
intermediate ages \citep{Kriek10}. Another problem related to each SSP
model is the convergence of the isochrones in the metal poor regime
which produces large uncertainties in the derived properties.

\begin{figure*}
\begin{center}
  \subfloat{\includegraphics[width=85mm,angle=270,trim=7mm 0mm 0mm 0mm,clip] {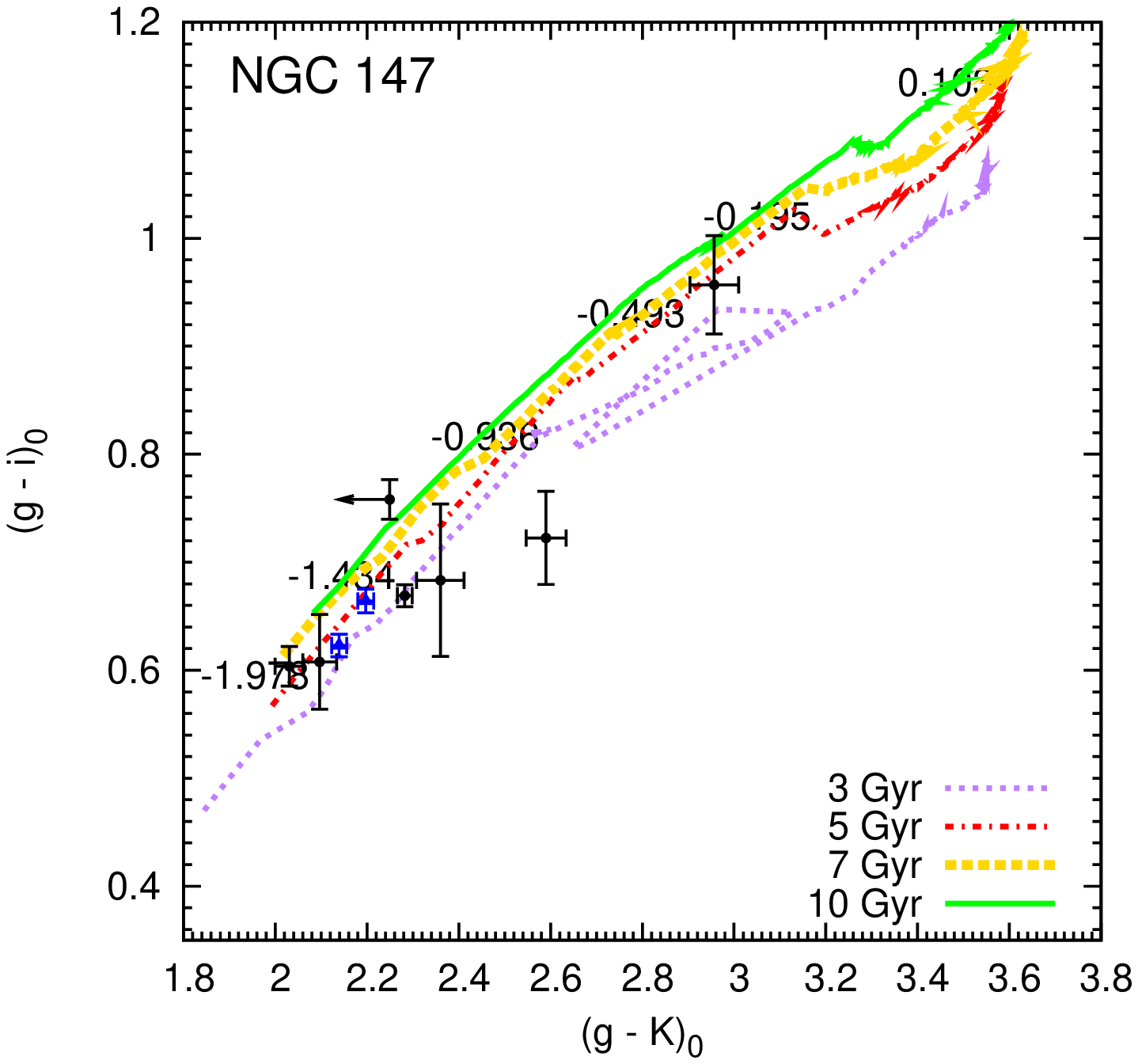}}          
  \subfloat{\includegraphics[width=85mm,angle=270,trim=7mm 0mm 0mm 0mm,clip] {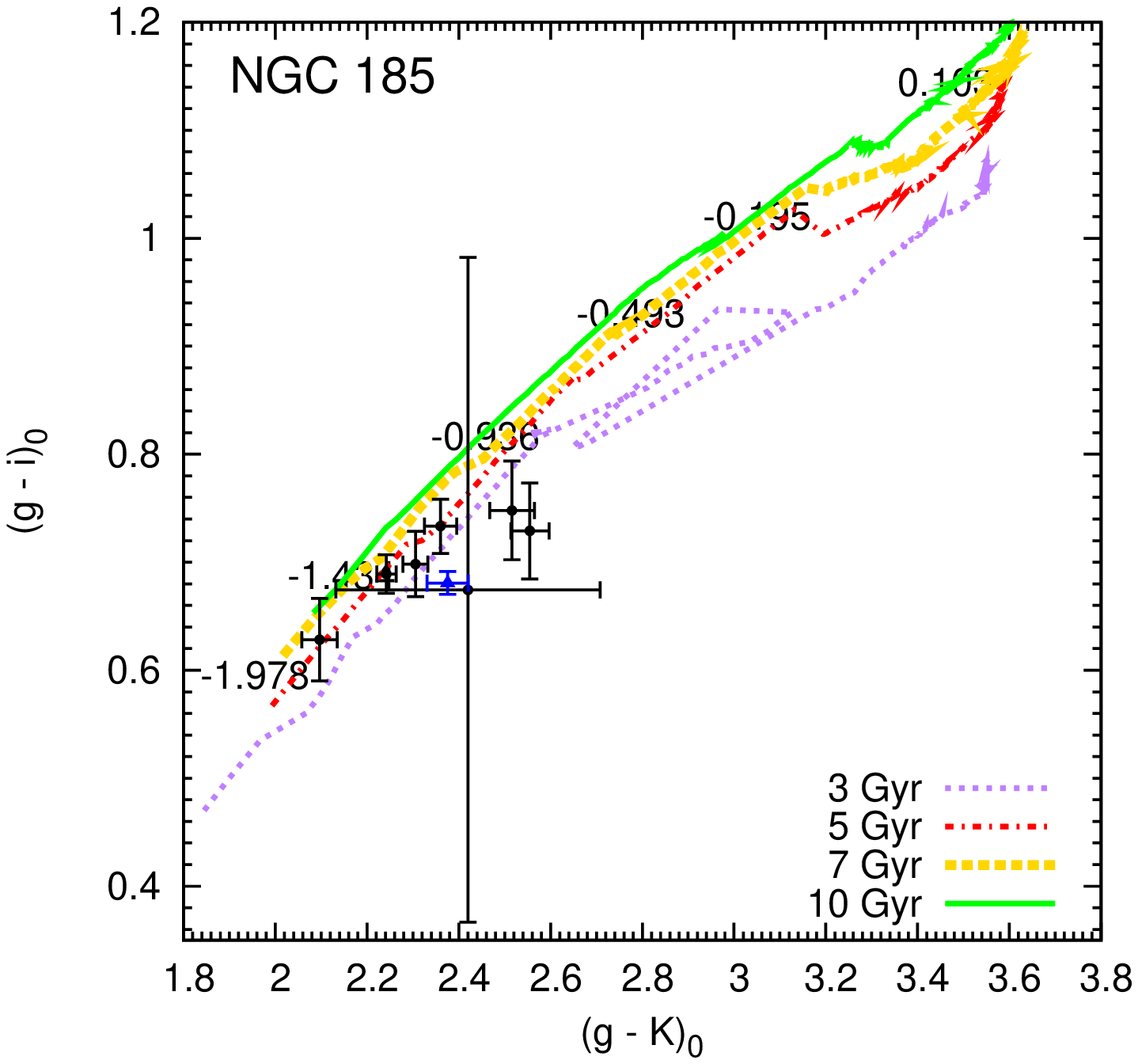}}
  \caption{($g-i)_0$ vs ($g-K)_0$ colours of GCs in NGC 147 (left) and
    NGC 185 (right) overlaid on top of isochrones from
    \citet{MarigoPadova08}. The corresponding metallicity ([Fe/H])
    values are indicated. The newly-discovered GCs are marked as blue 
    triangles. All but one cluster is found to be metal 
    poor, while the age is more difficult to constrain.}
\label{fig:SSP}
\end{center}
\end{figure*}

Figure \ref{fig:SSP} shows colour-colour diagrams constructed using 
the optical and near-IR photometry presented in Table 
\ref{tab:colors}.  While the SSP model track shown here is from 
\citet{MarigoPadova08}, we have checked that none of the results 
presented in this paper depend on the chosen SSP model. We resorted to 
using the native CFHT/Megacam rather than the more common 
Johnson/Cousins system in order to avoid uncertainties arising from the
colour transformations, as the SSP model tracks were readily available
for the CFHT/Megacam system. Figure \ref{fig:SSP} shows that all
clusters with the exception of Hodge IV in NGC 147 are metal poor with
[Fe/H]$\lesssim-1.25$ dex.  If Hodge IV is as metal-rich as we infer,
it would be rather interesting. However we note that it is the second
faintest cluster in NGC 147 and appears irregular in shape in the
MegaCam imaging (see Figure \ref{fig:cutouts}), while close to the
detection limit in the near-IR data. Although one of the first
clusters to be discovered in NGC 147 \citep{Hodge76}, high resolution
images of this object do not yet exist.  Even with high quality
ground-based data such as PAndAS, it is difficult to confirm it is a
genuine GC and not, for example, an asterism in the NGC 147 field.

The ages of the clusters are more difficult to estimate. At face
value, it appears that half of the GCs are old and half are of
intermediate age. However, inspecting the clusters that lie on the
young to intermediate age tracks, it is found that they are often
located in or near the main bodies of their host galaxies. As
discussed earlier, these regions are very crowded and often have
strong non-linear background gradients that can skew the colours of
the GCs.  This is certainly an issue for Hodge I in NGC 147 which has
a redder ($g-$K)$_0$ colour than any of the tracks shown.  The method
of photometric age estimation is not particularly accurate even when
free of these complications so we hesitate to claim we have found
evidence for the presence of intermediate age GCs.  Furthermore, due
to the way the SSP model isochrones are calibrated, small changes in
either of the colours used in Figure \ref{fig:SSP} will drastically
affect their age estimates, while only minimally impacting their
metallicity estimates.

Another outlier in Figure \ref{fig:SSP} is GC-SD10 in NGC 147 which
has a redder $(g-i)_0$ and bluer ($g-$K)$_0$ colour than any of the
tracks.  This object is the faintest out of all the clusters known in
the two dEs and was not detected in our K band imaging.  The quoted K
band magnitude represents an upper limit, suggesting that the actual
($g-$K)$_0$ colour must lie blueward of the point in in Figure
\ref{fig:SSP}. Although these colours seem unusual, GC-SD10 has been
spectroscopically confirmed by \citet{SharinaDavoust09} as a GC in NGC
147. Finally, even though the measured colours put FJJ III on the
model tracks on Figure \ref{fig:SSP}, the accompanying uncertainties
make constraining the age and metallicity of this cluster nearly
impossible.

An alternative way to determine the metallicities of GCs is to use
empirical colour-metallicity relations. We adopt the relationship
derived by \citet{KP02} (eq. \ref{eq:KP02}), calibrated using 129 M31
globulars that have $E$(B-V) $<$ 0.27, to further constrain the metal
content of the NGC 147 and NGC 185 GC systems. This relationship is
valid over the interval $-2.3 <$ [Fe/H] $< -0.2$~dex.

\begin{equation}
[Fe/H] = (V-K)\times1.82\pm0.11 - 5.52\pm0.26 
\label{eq:KP02}
\end{equation}
\begin{table}
 \centering
 \caption{ [Fe/H] values for GCs in NGC 147 and NGC 185. Columns refer to: (a) values obtained using the colour-metallicity relation from \citet{KP02};
   (b) \citet{daCostaMould88}; (c) \citet{sharina06ages,SharinaDavoust09}; (d) \citet{sharina06ages}. Superscripts refer to: (1) spectroscopic study; 
   (2) isochrone fitting. \citet{sharina06ages} originally presented [Z/H] values which were converted to [Fe/H] via the relation from \citet{Salaris06Book}.}
 \label{tab:cal}
 \begin{tabular}{@{}ccccc@{}}
 	\hline
	\hline
	ID			& [Fe/H](a)		& [Fe/H](b)$^1$		& [Fe/H](c)$^1$		& [Fe/H](d)$^2$ 		\\
				& [dex]			& [dex]			& [dex]			& [dex]				\\
	\hline
	Hodge I			&-1.3 $\pm$ 0.4 	& -1.9 $\pm$ 0.15	& -1.0 $\pm$ 0.5	& ...				\\
	Hodge II		&-1.7 $\pm$ 0.4 	& -2.5 $\pm$ 0.25	& -1.6 $\pm$ 0.5	& ...				\\
	Hodge III		&-2.1 $\pm$ 0.3 	& ...			& -1.8 $\pm$ 0.5	& -2.0 $\pm$ 0.1		\\
	Hodge IV		&-0.7 $\pm$ 0.4 	& ...			& ...			& ...				\\
	GC-SD5			&-2.2 $\pm$ 0.3	& ...			& -1.7 $\pm$ 0.3	& ...				\\
	GC-SD7			&-1.8 $\pm$ 0.3	& ...			& -1.6 $\pm$ 0.2	& ...				\\
	GC-SD10			&-1.9 $\pm$ 0.4 	& ...			& ...			& ...				\\
	PA-N147-1		&-2.0 $\pm$ 0.3 	& ...			& ...			& ...				\\
	PA-N147-2		&-1.9 $\pm$ 0.3	& ...			& ...			& ...				\\
	PA-N147-3		& ... 			& ... 			& ... 			& ... 				\\
	\hline
	FJJ I			&-1.3 $\pm$ 0.4 	& -1.4 $\pm$ 0.10	& -1.2 $\pm$ 0.4	& -1.6 $\pm$ 0.2		\\
	FJJ II			&-1.4 $\pm$ 0.4 	& -1.2 $\pm$ 0.25	& ...			& -2.1 $\pm$ 0.2		\\
	FJJ III			&-1.5 $\pm$ 0.5 	& -1.7 $\pm$ 0.15	& -1.4 $\pm$ 0.6	& -2.0 $\pm$ 0.1		\\
	FJJ IV			&-2.1 $\pm$ 0.3 	& -2.5 $\pm$ 0.25	& -1.6 $\pm$ 0.5	& -2.0 $\pm$ 0.2		\\
	FJJ V			&-1.8 $\pm$ 0.3 	& -1.8 $\pm$ 0.15	& -1.1 $\pm$ 0.6	& -1.5 $\pm$ 0.1		\\
	FJJ VII			&-1.7 $\pm$ 0.4 	& ...			& -0.4 $\pm$ 0.6	& ...				\\
	FJJ VIII		&-1.9 $\pm$ 0.3 	& ...			& -1.1 $\pm$ 0.9	& ...				\\
	PA-N185			&-1.6 $\pm$ 0.4 	& ...			& ...			& ...				\\

	\hline
 \end{tabular}
\end{table}
Table \ref{tab:cal} shows the metallicities derived with the above
relation and affirms our conclusion that the clusters are all indeed
metal poor. Although we do not use our spectra to derive
metallicities here, we note that the similarly strong Balmer lines
visible in Figure \ref{fig:sampleSpectra} for PA-N147-1 and
PA-N147-2 supports the similar metallicities derived from their
broadband colours. There is also a hint at a radial metallicity
trend with GCs with projected radii $>0.55$~kpc having systematically
lower [Fe/H] values by $\sim0.6$~dex compared to the central
population. For comparison, Table \ref{tab:cal} also shows
metallicities for some of the previously-known GCs derived from
spectroscopic studies
\citep{daCostaMould88,sharina06ages,SharinaDavoust09}, and from
colour-magnitude diagram isochrone fitting \citep{sharina06ages}.
There is generally a good agreement with past studies, lending further
confidence to our metallicity estimates. The only exception is FJJ VII
in NGC 185 for which a difference larger than 1 dex is seen between
our measurements and the spectroscopic measurements of
\citet{sharina06ages}. We currently have no explanation for this.

\subsection{Comparison to other Dwarf Elliptical Galaxies}

An oft-used statistic when comparing GC systems hosted by different
galaxies is the GC specific frequency, $S_{N} = N_{GC} \times 10^{0.4(
  M_{V} + 15 )}$, where $N_{GC}$ is the total number of GCs in the
system, and $M_{V}$ is the integrated absolute V magnitude of the host
galaxy \citep{Harris81}. This quantity can be thought of as the
formation efficiency of GCs relative to field stars, although the
interpretation is more complicated if one considers that fractions of
these populations may have been accreted rather than formed in situ.
In the case of dwarf galaxies, the globular cluster specific frequency
is poorly constrained at present due to incompleteness in terms of
imaging and GC detection, but it is critical for understanding what
kinds of dwarf galaxies may contribute GCs to the halos of more
massive systems.  Significant effort has been made to constrain this
value for dEs and past surveys have focused mainly on dwarf galaxies
that reside in dense environments, such as the Fornax and Virgo
clusters \citep{Durrell96,Miller98,Lotz04,Miller07,Peng08}. The study
by \citet{Miller07} found an overall trend of increasing $S_N$ with
increasing $M_V$ (decreasing galaxy luminosity), which has also been
found in dwarf irregulars although with smaller samples
\citep{Seth04,Georgiev08,Georgiev10}.  \citet{Peng08} find dwarf
galaxies with the highest $S_N$ values in their sample to be within 1
Mpc of the Virgo cluster core, which they interpret as an
environmental effect. However, these authors also find that dEs within
40 kpc of the most massive galaxies in the Virgo cluster have few or
no GCs, suggesting that they have probably been stripped away by the
tidal forces from the central hosts.

\begin{figure}
\begin{center}
  \subfloat{\includegraphics[width=80mm,angle=270,trim=0mm 0mm 0mm 0mm,clip] {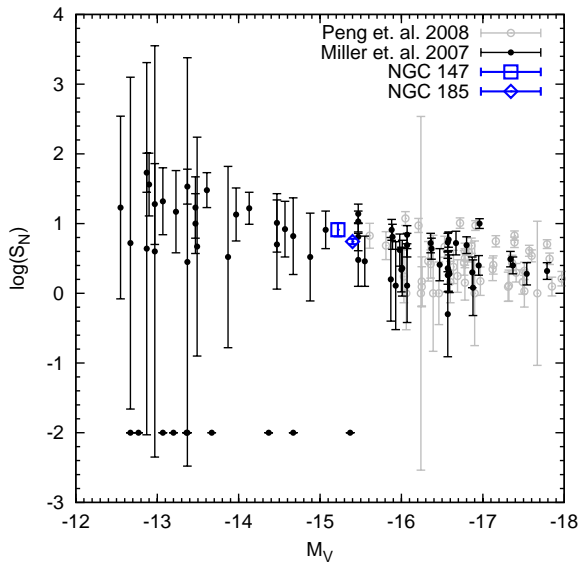}}          
  \caption{Plot of $log(S_N)$ vs. $M_V$ for dEs in the Fornax
      and Virgo clusters, and in the Leo group, taken from the studies
      by \citet{Miller07}(filled points) and \citet{Peng08}(open gray
      points).  Galaxies which do not host GCs are given $log(S_N) =
      -2$ in order to distinguish them on the plot. Overplotted are
      the corresponding $S_N$ values of NGC 147 and NGC 185.}
\label{fig:Sn}
\end{center}
\end{figure}

Using our updated GC census, the specific frequencies for NGC 147 and
NGC 185 were recalculated and found to be 8$\pm$ 2 and 5.5$\pm$0.5,
respectively. This is a slight increase from the previous values 
which were $6\pm2$ for NGC 147 and $4.8\pm0.5$ for NGC 185.  The
uncertainties quoted for NGC 147 allow for the possibility of Hodge IV
and SD-10 not being genuine GCs, as well as the uncertainty in the
galaxy luminosity. In the case of NGC 185, only the uncertainty in the
galaxy luminosity was taken into account.  Figure \ref{fig:Sn} shows
the $S_N$ versus $M_V$ values for dEs observed in the Fornax and Virgo
clusters and in the Leo group, taken from the studies of
\citet{Miller07} and \citet{Peng08}. The error bars are due to the
uncertainty in the host galaxy luminosity and the uncertainty in the
total number of GCs found around each galaxy, which has been corrected
for background contaminants and spatial incompleteness. On the same
figure overplotted are the updated $S_N$ values for NGC 147 and NGC
185. The $S_N$ values are within the range found for dEs of comparable
luminosity observed by \citet{Miller07}, albeit residing in different
environments, and appear to follow the trend of increasing $S_N$ value
with decreasing galaxy luminosity.

Another interesting property shared by the GC systems of NGC 147 and
NGC 185, and the GC systems of dE galaxies found in rich environments,
is their optical colours. The mean (V-I)$_0$ colours of NGC 147 and
NGC 185 GCs are 0.85 and 0.84 with standard deviations of 0.09 and
0.03 respectively. In their study of dEs in the Fornax and Virgo
galaxy clusters, \citet{Lotz04} have reported the peak of the mean
(V-I)$_0$ colour distribution to be 0.90 for galaxies with absolute
magnitudes between -15.0 and -16.0, and 0.85 for galaxies with
absolute magnitude between -14.0 and -15.0.  This makes the GC systems
of NGC 147 and NGC 185 nearly indistinguishable from those hosted by
similar luminosity dEs in rich clusters, suggesting that the
large-scale present day environment has little impact on either
$S_N$ or mean optical colour.

\subsection{Comparison to M31 Halo Globular Clusters}

Detailed surveys \citep[e.g.][]{Ferguson02,mcconnachie09,Richardson11}
have revealed complex substructure in the halo of M31, consisting of
loops, streams and filaments. This inhomogeneous halo is thought to
have formed over an extended period of time via the accretion and
disruption of dwarf galaxies.  M31 also hosts an extended population
of GCs, with clusters found up to distances of 200 kpc from its centre
\citep{mackey10a,Huxor11}. Evidence that the outer halo GC system may
have formed predominantly by accretion was hinted at by the radial
number density profile (see Figure 7 in \citet{Huxor11}). The profile
has the form of a broken power law, with characteristic flattening
occurring beyond a projected radius of $\sim$30 kpc. Such a behaviour
is found in the stellar halos of simulated galaxies that form via a
combination of accretion and in situ star formation \citep{Abadi06},
with the break in the power law marking the point beyond which the
bulk of the matter has been accreted.

\begin{figure}
\begin{center}
  \subfloat{\includegraphics[width=80mm,angle=270,trim=0mm 0mm 0mm 0mm,clip] {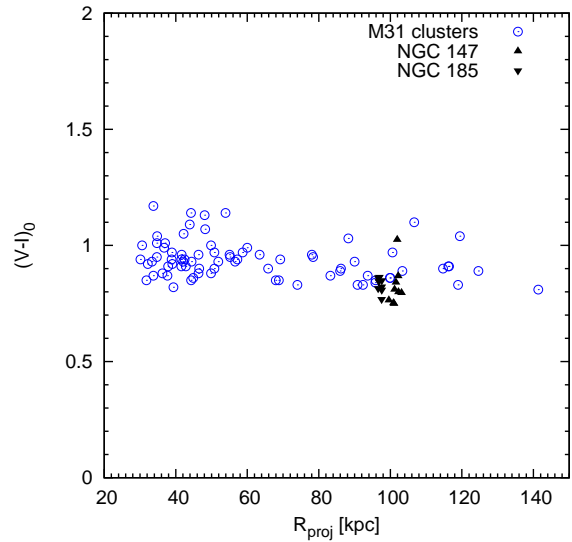}}          
  \caption{Plot of (V-I)$_0$ vs R$_{\rm proj}$ showing the M31 GCs
    that have projected radii larger than 30 kpc, together with the NGC 147
    and NGC 185 globular clusters. Data was taken from the Revised
    Bologna Catalogue \citep{GalletiRBC04} and Huxor et al. 2013 in
    prep. The colour similarity supports the suggestion that the M31
    outer halo GCs could have been accreted from objects like NGC 147
    and NGC 185. }
\label{fig:GC}
\end{center}
\end{figure}

More compelling evidence for an accretion origin comes from the fact
that GCs in the outer halo are spatially-correlated with the extended
substructure of M31 \citep{Mackey10b}. Statistical tests have shown
that GCs preferentially project on top of stellar streams, with the
probability of this being due to chance alignment being less than 1\%.
On this basis, it is argued that over 80\% of the GCs found beyond 30
kpc from the centre of M31 have probably been donated by captured and
disrupted dwarf galaxies.  The fact that there is a significant number
of dwarf galaxies in the halo of M31 \citep{Richardson11}, where a few
of the more luminous satellites are known to host GCs, shows that the
halo is still actively evolving.

If the majority of the M31 halo GCs were indeed donated by captured
dwarf galaxies, it might be expected that they share the some
properties with the GCs hosted by NGC 147 and NGC 185. Figure
\ref{fig:GC} displays the (V-I)$_0$ colours as a function of R$_{\rm
  proj}$ from the centre of M31 for the GCs hosted by NGC 147 and NGC
185. We also show confirmed M31 GCs that have R$_{\rm proj}$
\textgreater 30 kpc taken from the Revised Bologna Catalogue
\citep{GalletiRBC04} and Huxor et al. 2013 (in prep). The M31 GCs have
been de-reddened using extinction coefficients from
\citet{Schlegel98}.  One can see that the GCs found around the two
dwarf galaxies fit comfortably in the range of (V-I)$_0$ colours
measured for their M31 halo counterparts. Although this does not prove
that the M31 halo GCs have been accreted by systems like NGC 147 and
NGC 185, it demonstrates that this idea is not inconsistent with the
data. In terms of the number of GCs they currently host, only about 8
NGC 147 or 185 systems would be needed to create almost the entire M31
halo GC population that consists of $\sim$80 clusters.  Such
accretions would also donate $\sim 5.4\times10^8$~L$_{\odot}$ of
stellar light, consistent with estimates of the halo luminosity
\citep{Irwin05}.

\section{Summary}

We have presented the results of a new search for remote GCs around
the M31 satellites, NGC 147 and NGC 185, using data from PAndAS. The
search resulted in the discovery of four new GCs, three of which are
located close to NGC 147, while one is near NGC 185.  Probability
arguments and newly obtained radial velocities indicate that these
objects are likely to belong to the dwarfs and not the M31 halo. Our
findings serve as another example of the importance of studying
galaxies beyond their optical boundaries.

We present the first homogeneous optical and near-IR photometry for
the entire GC systems of these two dwarf galaxies. We use this to
constrain the GC ages and metallicities via the use of colour-colour
plots and empirical colour-metallicity relations, finding that, in
general, the clusters are old and metal poor ([Fe/H]$\lesssim-1.25$
dex).

The mean colours of the GCs hosted by NGC 147 and NGC 185 are found to
lie at the peak of the colour distribution of the GC systems belonging
to dEs in the Fornax and Virgo galaxy clusters, despite a large
difference in the environments in which they reside. Their $S_N$
values are consistent with the trend of increasing GC specific
frequency with decreasing galaxy luminosity generally observed for
dwarf galaxies, regardless of their type, and in a variety of
environments. The close similarity between the (V-I)$_0$ colours of
the GCs belonging to these two dwarf galaxies and those belonging to
the M31 outer halo is consistent with the idea that accretion of the
former could have contributed to the assembly of the latter.

\section*{Acknowledgments}

The work of APH was partially supported by Sonderforschungsbereich SFB
881 ``The Milky Way System" of the German Research Foundation (DFG).
ADM is grateful for support from the Australian Research Council
through an Australian Research Fellowship (Discovery Projects grant:
DP1093431). We thank the CFHT staff for their support and helpfulness
throughout the PAndAs project.  Based on observations obtained with
MegaPrime/MegaCam, a joint project of CFHT and CEA/DAPNIA, at the
Canada-France-Hawaii Telescope (CFHT) which is operated by the
National Research Council (NRC) of Canada, the Institut National des
Science de l'Univers of the Centre National de la Recherche
Scientifique (CNRS) of France, and the University of Hawaii. This work
is based in part on data products produced at TERAPIX and the Canadian
Astronomy Data Centre as part of the Canada-France-Hawaii Telescope
Legacy Survey, a collaborative project of NRC and CNRS.  Based on
observations made with the 4.2m William Herschel Telescope operated on
the island of La Palma by the Isaac Newton Group in the Spanish
Observatorio de Roque de los Muchachos of the Instituto de Astrofisica
de Canarias.  The United Kingdom Infrared Telescope is operated by the
Joint Astronomy Centre on behalf of the Science and Technology
Facilities Council of the U.K.


\appendix
\chapter{}
\label{appendix:a}

\section[]{Review of the Literature on the NGC 147 and NGC 185 GC
  Systems}

The following is a short review of the discovery history and the
nomenclature of the GCs hosted by NGC 147 and NGC 185. The motivation
for this is to highlight some inconsistencies in the literature that
we discovered while this paper was in preparation. By doing so, we
hope to minimise the possibility of future confusion when studying the
GCs of these two dEs.

The existence of GCs in both NGC 147 and NGC 185 was first reported by
\citet{Baade44}, who discovered two globulars in each of the galaxies.
In his paper, the clusters were not named and coordinates were
provided only for the ones hosted by NGC 185 and in terms of relative
positions (measured on photographic plates) from the galaxy centre.

\citet{Hodge74} reported the discovery and presented photometry of
five GCs in NGC 185, two of which were those previously discovered by
\citet{Baade44}. These clusters were simply labelled 1-5. While their
coordinates were not provided, a finding chart was shown.

Two years later, \citet{Hodge76} published a paper on the structure of
NGC 147, in which the discovery and photometry of two additional GCs
bound to this galaxy were presented, alongside the two clusters
previously discovered by \citet{Baade44}. Once again, no coordinates
for any of these objects were given, but a finding chart was
published, on which the clusters are labelled 1-4. In the literature,
these clusters are now known as Hodge I-IV.

In an Appendix to their paper on planetary nebulae in NGC 147 and NGC
185, \citet{FJJ77} revisited the GC systems of the two dEs. In
addition to the GCs already discovered by Baade and Hodge around NGC
185, they presented the discovery of an additional four clusters from
their photographic plates, while also discarding the object labelled
by \citet{Hodge74} as ``2'' as a GC. In \citet{FJJ77}, the objects are
numbered I-VIII but the counting does not follow the pattern started
by \citet{Hodge74}. This nomenclature has propagated through the
literature, and these clusters are referred to as FJJ I-VIII in
recent publications. In addition, equatorial coordinates together with
a finding chart were also published for the entire sample of GCs
described by \citet{FJJ77}.

\citet{FJJ77} also revisited the GC system of NGC 147. They recovered
the objects already identified by \citet{Hodge76} as globulars and did
not find any new members belonging to this system. They showed a
finding chart and a table with equatorial coordinates for the three
brightest globulars in this galaxy. Their finding chart is identical
to the one published by \citet{Hodge76} in terms of the labelling and
positions of the GCs: Hodge II is south of Hodge I and Hodge III 
is south of Hodge II. However, in their Table 9 that lists
the clusters' coordinates, the positions of Hodge II and Hodge
III are swapped: Hodge II is listed to be south of Hodge
III. This unintentional oversight is most probably the main reason
for many inconsistencies in the more recent literature.

A paper published by \citet{daCostaMould88} presented spectroscopic
data and metal abundances of the GCs hosted by NGC 147 and NGC 185.
Regarding the NGC 185 system, they presented V-band photometry taken
from \citet{Hodge74} and spectroscopic data for clusters FJJ I-V,
as well as Hodge 2 that \citet{FJJ77} classified not to be a
cluster.  For cluster FJJ IV which was not listed amongst the GC
candidates by \citet{Hodge74} the V magnitude was estimated by eye.
Analysing the cluster spectra, \citet{daCostaMould88} showed that
Hodge 2 is indeed a galaxy at redshift $z=0.04$ and not a GC.

Regarding the NGC 147 system, \citet{daCostaMould88} took spectra only
of Hodge I and Hodge III. In their Table 1 they listed the
photometric V magnitude of these two clusters as reported by
\citet{Hodge76}. They did not list coordinates for any of the clusters
but stated that the centres of the clusters were taken from
\citet{FJJ77}.  This probably means that they presented metal
abundances of Hodge II rather than Hodge III.  There was no
indication that \citet{daCostaMould88} noticed the oversight made by
\citet{FJJ77}, and in Table \ref{tab:cal} in our paper it is assumed
that they have not.

\citet{Geisler99} observed all but one cluster in NGC 185 with the
HST, that were known at that time. The cluster which was not 
observed was FJJ VIII. They found that FJJ VI is not a GC but an 
elliptical galaxy. All other cluster candidates that were observed 
were confirmed to be genuine GCs.

In more recent literature, \citet{sharina06ages} revisited the GC
systems of NGC 147 and NGC 185 using HST/WFPC2 imagery and
spectroscopy taken with the SCORPIO spectrograph. They did not provide
coordinates for any of the clusters, but they did provide finding
charts. In the case of NGC 147, they had the positions of Hodge II
and Hodge III reversed compared to the original publication by
\citet{Hodge76}, so it is highly likely the data and results presented
for Hodge II actually refer to Hodge III and vice versa.  In
their Table 2, they have taken the V magnitudes of what they label as
Hodge I and Hodge III from the original paper by \citet{Hodge76}, 
while the V magnitude of their Hodge II is taken
from \citet{Hodge74} even though this paper analysed only clusters
hosted by NGC 185 and did not list photometric values for any cluster
in NGC 147. This made the cluster they label as Hodge II brighter
than Hodge III, which is easily seen not to be the case with
simple visual inspection of their HST images.

The most recent publication regarding the NGC 147 GC system is by
\citet{SharinaDavoust09}. In their paper they announced the discovery
of three new GCs. They presented a coordinate table and a finding
chart, both having correct positions compared to the original
\citet{Hodge76} publication regarding the ``classical Hodge''
clusters.  However, it was not stated that there are differences in
the positions compared to \citet{sharina06ages}. It is also possible
that any values they cited from the \citet{sharina06ages} paper might
be assigned to the wrong object.

Finally, in our paper we report the discovery of another three
GCs hosted by NGC 147, and one hosted by NGC 185. Interestingly, all three of the new
GCs in NGC 147 lie beyond areas previously imaged for GC searches, while
the new GC in NGC 185 lies within the photographic plate region searched by \citet{FJJ77}.
We speculate that the outlying nature of this object coupled with its low luminosity
caused it to be missed in the original study. The tables we
present here reflect the original naming and correct coordinates for all
previously-known GCs, as well as these new discoveries.

\bsp

\label{lastpage}

\end{document}